\documentstyle[11pt]{article}

\begin{document}

\begin{center}

GALAXY ASSOCIATIONS WITHIN THE COMA CLUSTER
\vspace{0.2in}

V.G.Gurzadyan$^{1}$ and A.Mazure$^2$
\vspace{0.1in}

1. ICRA, Dipartimento di Fisica, Universita di Roma La Sapienza,
   Rome, Italy and Yerevan Physics Institute, Yerevan, Armenia

gurzadya@icra.it

2. Laboratoire d'Astronomie de Marseille, Marseille, France

alain.mazure@astrsp-mrs.fr
\vspace{0.2in}

\end{center}

{\bf Abstract}

The mean redshift of the core of the Coma
cluster - $cz = 6953$ km s$^{-1}$ and its dispersion
$\sigma=949$ km s$^{-1}$
are obtained by means of the analysis of the substructures
of this
cluster by using the S-tree method.
The existence of three subgroups of galaxies is revealed,
one of them is associated with the cD galaxy NGC 4874,
the other -- with NGC 4889.
It is argued that these subgroups are {\it galaxy
associations}, i.e.
galactic dynamical entities moving within the main
cluster. Thus we conclude
that the non-stationarity of the dynamical processes
ongoing in the Coma core
is due to the merging of small-scale galaxy associations,
rather than of
two equal sized clusters.
We provide the lists of the galaxies of the associations, the
observational study of which can be of particular interest.

\vspace{0.1in}



\section{Introduction}

In (Gurzadyan \& Mazure, 1998) we have
studied the substructure
of several Abell clusters from the ESO Key Program ENACS
database (Mazure et al
1996). The analysis was carried on using the S-tree
method (Gurzadyan
et al 1994).
It was shown that the studied clusters contain 2-3
subgroups, denoted, due
to their remarkable dynamical properties, as {\it galaxy
associations}. Those
dynamical properties include the truncated
1D velocity distribution and relatively stable velocity
dispersion.
It was argued that
the motion within the main cluster can be the natural
mechanism explaining
such a truncation. Because of the essential role that
galaxy associations can have for the understanding of the
evolution of
galaxy clusters, their further study seems of particular
importance.
This concerns especially observational studies, since the
galaxies
of the associations have to exhibit anomaly strong star formation
activity,
anomalies of a disk and bulge for spirals, possible
tidal tails, etc, with
more probability that ordinary galaxies of the cluster.

Here we analyze the core substructure
of the Coma cluster using the S-tree method.
The Coma cluster is one of the most studied clusters
with remarkable substructure
(e.g. Rood \& Sastry, 1971; Valtonen \& Byrd, 1979; Baier
1984;
Mellier et al 1988; Dressler \& Shechtman 1988; Escalera et al 1992;
Fabian et al 1994; Gambera et al 1997;
Kashikawa et al 1998); for more references see (Mazure, et al
1998).
The internal dynamics and the substructure of Coma have
been
extensively analyzed by Biviano et al (1996);
Colless \& Dunn, (1996). Though the main conclusions of
both studies are in agreement, there are some differences
concerning
the structure of the central region of the cluster.
In general, the main tools in such investigations,  on
one hand, involve
various statistical methods, the wavelets being among the
efficient ones, and
creation of various models based on certain assumptions
on the cluster's
symmetry, equilibrium, dark matter distribution, etc.
(e.g. Merrit 1987), on the other hand.

The X-ray observations of the Coma cluster, especially by
ROSAT (White et al 1993; Vikhlinin et al 1997), add
crucial information
on the substructures, as well as on the processes
governing the
mechanisms of X-ray emission.
Both the galactic data and the X-ray images of Coma
support the
existence of substructures in the cluster, and the
problem comes
to the clear identification
of the basic dynamical mechanisms
responsible for
the evolution of the cluster.
Clearly, the further combination of data on the
galaxies and X-ray data will enable much deeper insight
onto the structure
of the cluster, though one has to better understand the
dynamics
within the cluster, in order to proceed from realistic
assumptions on
the equilibrium conditions on the X-ray emitting gas.

Our study of the Coma core reveals the existence
of 3 subgroups, which appear to be galaxy associations.
Two of the latter are associated with the brightest cD
galaxies
NGC 4874 and NGC 4889, respectively. Their dynamical
parameters
are established, which enables us
to conclude that they are moving within the cluster, so
that one is a
young merger, while the subgroup of NGC 4889 has been
already  largely dissolved.
The segregation of galaxies within the subgroup we
incline to explain not by core-halo evolution,
but via dynamical friction during the motion.
The obtained results indicate this essentially
non-stationary
character of structure of Coma cluster, and do not
support the view
that the latter can be a result of merging of two
clusters of equal size.

\section{Data}

In the analysis below we have used the data compiled by
Biviano et al (1996); their dataset of galaxy redshifts
is based
on their own observations with CFHT, on the data by
Colless and Dunn (1996)
and those available in the literature. The authors
provide also the
completeness data with respect to the limiting magnitude.
From this dataset we have extracted a sample of 188
galaxies in a  $3000''\times 3000''$ field centered
on $\alpha=12^h 57^m.3$, $\delta=28^{\circ} 14'.4$ and
brighter than $18^m.0$ (Mazure and Gurzadyan 1998).
This choice was determined by data completeness with respect to
magnitude.

\section{Method}

The S-tree technique developed for the study of the hierarchical
substructure
of galaxy clusters is described in detail in (Gurzadyan et al 1994;
Gurzadyan \&
Kocharyan 1994).
It is based on the
methods of theory of dynamical
systems and  has already been used for the  study of
substructure of
the various clusters of galaxies, including the
Local Group,
Virgo (Petrosian et al 1998) and  the sample of above
mentioned
ENACS Abell clusters.

This method is using the information on the
2D coordinates, redshifts and magnitudes of galaxies in a
self-consistent
way, namely, revealing the correlation which should exist
between the
particles' parameters (coordinates and velocities) of a
gravitationally
interacting N-body system. It is reached using a well
know method
in classical mechanics reduced to the study of the
properties of the
flow of geodesics in phase space of the system (Arnold,
1989). Namely
the so-called two-dimensional curvature
$$
K_{\mu\nu}=Riem_{\rho\delta\mu\nu}u^{\rho}u^{\delta}
$$
($Riem$ is the Riemannian
tensor, $u$ is the velocity of geodesics) is used for the
evaluation of the 'degree
of boundness'; for details see (Gurzadyan \&
Kocharyan 1994) and the Appendix in (Gurzadyan \& Mazure,
1998).
This procedure enables to reveal the
structure of the system including the
existence of subgroups
and representation of the result via
tree-diagrams (S-tree). In the above quoted papers the role of
the magnitude completeness, M/L ratio of galaxies and other
effects
are discussed as well. Special algorithms are used also
for the
analysis of the outcome information (Bekarian \&
Melkonian 1997).

\section{Substructure of the Coma Core}

This is not the first cluster studied via the S-tree method, therefore we
will not
repeat the description of the technical steps of the analysis.
Thus, our analysis revealed the substructure of the core of Coma cluster,
namely, the complex structural and
dynamical conditions in the core of Coma cluster.
First, the code has enabled to extract the galaxies with correlated
parameters, thus defining
the main physical system (MS) of galaxies: it contains 174
galaxies, centered at $\alpha=12^h 57^m 32^s.3$,  $\delta=28^{\circ}
19'31^"$.
The knowledge of the membership of galaxies is readily defining the
redshift of the Coma cluster and velocity dispersion:
$cz = 6953$
km s$^{-1}$ and
$\sigma=949$ km s$^{-1}$, respectively.

Then, at higher level of correlation, i.e degree of the mutual boundness,
the existence
of 3 subgroups of MS has been revealed. The subgroups contain 34 galaxies
(1s), 14 galaxies (2s) and 17
galaxies (3s), as exhibited in Tables 1-3.

The 1st subgroup includes the 2nd brightest galaxy of the
Coma core - NGC
4874. The center of this subgroup lies at $\alpha=12^h
57^m 34^s.31$,
$\delta=28^{\circ} 15' 35^".45$, i.e. does not coincide
with NGC 4874.

The obtained parameters of galaxies of the MS and the
subgroups are presented
in  Table 4, which includes the number of galaxies
(N), the median
velocity (m, in km s$^{-1})$), standard deviation of the
redshift
distribution ($\sigma$, in km s$^{-1}$),
3rd and 4th moment of the redshift distribution, (s) and (c),
respectively. We do not include any estimation of the error box
for the standard deviation, since its precise value is not only of
minor importance for our main aim of the subgrouping,
but also due to the inhomogeneity of the input data, any weighting,
strictly speaking, will add a bias in such estimation.

Figure 1 shows the histograms of the redshift
distributions of the initial
sample, the MS, and of the three subgroups.

\section{Discussion}

The results obtained above
enable us to draw the following picture on the
substructure and the dynamical
processes evolving in the Coma core.

 We have identified the galaxies forming the main body of the Coma
cluster and have obtained its redshift and the velocity
dispersion: $cz = 6953$ km s$^{-1}$ and
$\sigma=949$ km s$^{-1}$,
respectively. These parameters do not differ much from
those obtained before
(Biviano et al, 1996, Colless \& Dunn 1996), except the
center of the cluster
does not coincide exactly with the dominant galaxy NGC
4874, though lies in
its vicinity.

Then, we have revealed 3 subgroups of galaxies within the main system and
have determined the charactersitics of each of them, given in Table 4.
Careful look of those parameters enables to understand certain dynamical
processes going on within the cluster. Namely,  the clear separation of
the subgroups in the redshift space
indicates their essential mutual bulk motion (Gurzadyan \& Mazure
1998), i.e. when
the bulk velocity is exceeding the velocity dispersion of
each subgroup.
This can mean ongoing merging of the
subgroups, as confirmed below by additional arguments.
The 3rd and 4th momentum show the
absence of asymmetry on one hand, and stronger truncation of subgroups'
(as compared with
the main system) redshift distribution, on the other
hand. Truncation of the subgroups, i.e. the cutoff of the
the velocity dispersion of galaxies in the subgroups must occur
at the motion of the subgroups within the main system (Gurzadyan and
Mazure 1998).

Further evidences support this conclusion.  NGC 4874 ($\alpha=12^h 57^m
27^s.38$,
$\delta=28^{\circ} 13' 43^"$, $cz=7176$ km s$^{-1}$)
is not situated at the mass center of subgroup 1
and its redshift does not coincide with the median
redshift of the subgroup containing the several bright galaxies of the
Coma cluster, i.e. 5 galaxies brighter 15$^m$ (see Table 2).
Such
a redistribution of galaxies is inevitable due to dynamical friction,
if the group of galaxies is moving through a galaxy field.

Biviano et al (1996) proposed to explain the overdensity
of bright galaxies
in the vicinity of NGC 4874 by the core-halo segregation
mechanism
during the own evolution of the subgroup. However, this
mechanism
is efficient in isolated stellar systems with large
number of stars, while at least the required virialization
is never reachable for a small-number galaxy group moving within a 
giant host system. 
The dynamical friction, on the other hand, does not depend
on the smallness of the number of galaxies since acts on each galaxy
individually, depending on the mass and the velocity of the moving object,
and hence can be responsible for the observed segregation; for the
observed 
galaxy/subgroup parameters the efficiency of dynamical friction requires
time scales $10^8 - 10^{10}$ yrs, i.e. cosmologically quite reasonable
scale. This
is again supporting the significance of the regular motion of the subgroup
within the host cluster.
Subgroup 1 shows better separation by 'degree of
boundness' from the
galaxies of the main cluster in redshift space, while the
galaxies of subgroup
3 are more overlapped with redshifts of main cluster,
i.e. there are galaxies
not belonging to subgroup 3, but having redshifts lying
within the
redshift interval of that subgroup.
Since one can hardly accept the possibility of the formation of the
dynamical subgroups during the evolution of the main cluster and hence
doubt
in their primary origin and in their further dissolution within the
cluster,
the above fact can  indicate only the more stronger dissolution of
subgroup 3,
as compared with the subgroup 2.
In other words the former has to be an elder merger; this conclusion
is supported also with the
essential shift of the velocity of NGC 4889
($\alpha=12^h 56^m 55^s$,
$\delta=28^{\circ} 14' 46^"$, $cz=6497$ km s$^{-1}$)
from the median
velocity of subgroup 3, a result of an action of dynamical friction.
Similar conclusion has been drawn by Colless and Dunn
(1996) from other
considerations.

Thus, the present study indicates the existence of
subgroups - galaxy associations  -
in Coma cluster which are undergoing the merging process, since
essential bulk velocities and dissolution must mean only merging
and not any time-reversed processed.
It is remarkable that the subgroups are
in various phases of merging. Indeed, the truncation
of redshift
distribution for the elder merger
(subgroup 3) is more evident than for subgroup 1,
so that only
a core of galaxies had survived in the subgroup
3.
Essential bulk velocity has been revealed
also for the
subgroups of the Local group  (Gurzadyan \& Rauzy, 1997;
Rauzy \& Gurzadyan, 1998).

The existence of merging subgroups with
mutual bulk
motion within the host cluster does not
support the idea that
the Coma cluster is a result of merging of two equal
sized clusters
(e.g. Tribble 1993) but is in agreement with one of the
alternatives mentioned
already by Fitchett and Webster (1987).

The fact that   essentially non-stationary
dynamical processes are ongoing
in the core of Coma cluster  affects the
interpretation of the X-ray
data as a way to reveal the structure of the cluster;
see (Mazure et al 1998; Kikuchi et al 2000) for  references.
Particularly, the isothermal assumption on the
X-ray gas state  leads to an overestimation of
the mass of the system within the given radius.
A multi-temperature gas will mean that the
X-ray flux peaks
will depend on the wavelength, and therefore the
correlation
of the X-ray peaks with galaxy distributions may be
not straightforward (cf. Biviano et al 1996) .

The available data on the galaxies of the subgroups given in the Tables
1-3
show some peculiarities. For example, the subgroup 3 is almost totally
composed
of S0 galaxies. Seems remarkable also the non-random orientation
of the galaxies in subgroup 1 (Fig. 2).

Another observational aspect concerns the individual properties of  the
galaxies of the subgroups.
It is now becoming clear that even minor external perturbations can be the
reason of various galactic anomalies,
such as the starburst activity (Bekki 1999), disk thickening of spirals
(e.g. Reshetnikov and Combes 1997),  bar formation
and misalignment (Curir and Mazzei 1999, Chitre et al 1999), low
metalicity globular cluster formation (Tanighuci et al 1999, Smith 1999),
tidal tails and bridges (Conselice and Galagher 1998), counter rotating
disks, etc.  The galaxies of subgroups with essential
bulk velocity have to undergo more perturbations and hence to reveal more
of those properties than the
galaxies of the field. Hence the observational study of the galaxies
listed in Table 1-3 can be of particular interest, since the
above mentioned tidally triggered phenomena link
the dynamically detected galaxies with their individual
properties.

We are thankful to F.Combes and G.Comte for valuable discussions.
V.G. was supported by French-Armenian Jumelage.

\begin{table}             
\caption{Subgroup 1.  The columns indicate: Col 1 GMP No,
 Col 2 Velocity from Biviano et al (1996),
 Col 3 and  4 X, Y in arsec from GMP,
 Col 5 B mag from GMP,
 Col 6 R mag from GMP,
 Col 7 B-R from GMP,
 Col 8  Ellipticity from GMP,
 Col 9 Orientation from GMP,
 col 10 V mag from GMP and GP,
 Col 11 Morphological type from Biviano et al compilation:
 (Morphological type from Dressler(1980),
           or from M88 (2nd choice), numerically coded as follows:
           E=1.0, S0=2.0, Sa=3.0, Sb=4.0, generic SP=5.0,
           Sc=6.0, Sd=7.0, Sm=8.0, I=9.0.
           Intermediate or uncertain cases, such as "E/S0" or "S/I"
           have a number intermediate between the two classes, e.g.
           E/S0=1.5, S/I=8.5.
           Barred galaxies have their number incremented by 0.1, e.g.
           S0B=2.1, SBb=4.1.
           Peculiar galaxies have their number incremented by 0.2,
           e.g. Sp=5.2, S0p=2.2).
 Col 12:   Spectrum classification, according to C93:
          0=normal; 1=abnormal, with some doubts; 2=abnormal.
 Col 13:   Galaxy number in IC/NGC catalogues.
}
{
\renewcommand{\baselinestretch}{1.2}
\renewcommand{\tabcolsep}{.95mm}
{\footnotesize
\begin{center}
\begin{tabular}{ccrrcrccrcccl}
\hline
\hline\\[-3.5mm]
GMP& VEL& \multicolumn{1}{c}{X}& \multicolumn{1}{c}{\,\,\,\,Y}& B\,&
\multicolumn{1}{c}{\,\,R}& B-R & Ell & The & V\,& Type & Class & Ident\\
\hline\\[-3mm]
2615& 6707& -627& -736& 16.97&  0.00& 1.90&  0.3&  98& 15.70& 2.5& 0&
\\
2535& 7056& -736&   94& 15.93&  0.00& 1.90&  0.3&  49& 14.75& 2.0& 0&
IC4041  \\
2654& 6984& -565&  -52& 16.38&  0.00& 1.90&  0.3&  13& 15.21& 2.0& 0&
\\
2798& 6811& -429&  -54& 14.85&  0.00& 0.00&  0.0&   0& 13.82& 1.0& 0&
NGC4898A\\
2866& 6992& -363& -680& 16.90&  0.00& 1.79&  0.2&  18& 15.73& 1.0& 0&
\\
2912& 6756& -316&  719& 16.07&  0.00& 1.80&  0.4&  96& 14.93& 1.0& 0&
\\
2922& 7196& -303&  388& 15.93&  0.00& 1.86&  0.1& 149& 14.75& 1.0& 0&
IC4012  \\
2940& 7245& -280&  120& 16.08&  0.00& 1.82&  0.1& 149& 15.24& 1.0& 0&
IC4011  \\
2943& 7249& -279& 1008& 17.66& 16.39& 0.00&  0.3& 163& 16.73& 0.0& 0&
\\
3017& 6784& -209&  -91& 17.91& 16.28& 1.65&  0.2& 104& 17.08& 1.0& 0&
\\
3055& 6691& -168&  994& 14.73&  0.00& 1.87&  0.1& 152& 13.60& 1.0& 0&
NGC4881 \\
3129& 6729&  -69&  626& 17.94& 16.26& 1.71&  0.5& 173& 16.78& 0.0& 0&
\\
3201& 6629&   10& -210& 15.51&  0.00& 1.91&  0.0&   0& 14.46& 1.0& 0&
NGC4876 \\
3206& 6892&   14&  -43& 16.36&  0.00& 1.79&  0.4&  18& 15.81& 2.0& 0&
\\
3213& 6841&   19&   87& 16.14&  0.00& 1.83&  0.1& 131& 15.21& 1.0& 0&
\\
3222& 6946&   38& -166& 16.47&  0.00& 1.75&  0.1& 162& 15.53& 1.0& 0&
\\
3238& 6722&   53&-1119& 16.75&  0.00& 1.88&  0.2& 124& 15.55& 4.0& 0&
\\
3291& 6812&   92&   61& 16.41&  0.00& 1.78&  0.3&  45& 15.73& 2.0& 0&
\\
3329& 7176&  124&  -41& 12.78&  0.00& 0.00&  0.0&   0& 12.20& 1.0& 0&
NGC4874 \\
3352& 7205&  147&  -86& 14.79&  0.00& 1.78&  0.0&   0& 14.29& 1.5& 0&
NGC4872 \\
3390& 6832&  183&  275& 15.89&  0.00& 1.75&  0.3& 120& 14.33& 2.0& 0&
\\
3414& 6717&  202&  -52& 14.89&  0.00& 1.90&  0.4& 173& 14.19& 2.0& 0&
NGC4871 \\
3423& 6817&  211& -434& 15.80&  0.00& 1.95&  0.5& 154& 14.65& 2.0& 0&
IC3976  \\
3471& 6665&  247&  101& 16.45&  0.00& 0.00&  0.2& 148& 15.76& 1.5& 0&
\\
3510& 6992&  289& -214& 14.97&  0.00& 2.06&  0.1& 173& 13.85& 1.0& 0&
NGC4869 \\
3554& 7125&  332&  374& 17.20& 15.33& 1.87&  0.1& 140& 16.03& 2.0& 0&
\\
3660& 6729&  422& -704& 15.76&  0.00& 1.87&  0.2&  86& 14.66& 2.0& 0&
IC3963  \\
3664& 6806&  426&   24& 14.70&  0.00& 0.00&  0.0&   0& 14.19& 1.0& 0&
NGC4864 \\
3707& 7220&  472&  255& 17.76& 15.96& 1.82&  0.4& 116& 16.56& 2.0& 0&
\\
3706& 6892&  472& -371& 17.61&  0.00& 1.85&  0.1& 167& 16.44& 1.0& 0&
\\
3730& 7053&  492& -670& 15.27&  0.00& 1.94&  0.1&  26& 14.16& 1.0& 0&
IC3959  \\
3794& 6960&  546&  -40& 17.37&  0.00& 1.98&  0.1& 101& 16.12& 1.0& 0&
\\
3896& 7378&  646& -491& 15.13&  0.00& 1.75&  0.0&   0& 14.18& 5.0& 2&
IC3949  \\
4230& 7198&  992&  162& 15.19&  0.00& 1.87&  0.1&  86& 14.03& 1.0& 0&
\\
\hline
\end{tabular}
\end{center}
}}
\end{table}

\begin{table}                 
\caption{Subgroup 2.}
{
\renewcommand{\baselinestretch}{1.2}
\renewcommand{\tabcolsep}{.95mm}
{\footnotesize
\begin{center}
\begin{tabular}{ccrrcrccrcccl}
\hline
\hline\\[-3.5mm]
GMP& VEL& \multicolumn{1}{c}{X}& \multicolumn{1}{c}{\,\,\,\,Y}& B\,&
\multicolumn{1}{c}{\,\,R}& B-R & Ell & The & V\,& Type & Class & Ident\\
\hline\\[-3mm]
2541& 7494& -722&  -168& 15.44&  0.00& 1.98& 0.0&   0& 14.29& 1.0& 0&
NGC4906\\
2559& 7627&   97&  -695& 15.44&  0.00& 0.00& 0.6& 152& 14.83& 6.5& 0&
IC4040 \\
2551& 7537& -709&   158& 16.85&  0.00& 2.99& 0.3&  78& 15.36& 2.1& 1&
\\
2651& 7679& -571&     6& 16.19&  0.00& 1.85& 0.3& 101& 15.07& 2.0& 0&
\\
2721& 7579& -493& -1249& 17.50& 15.65& 1.82& 0.4& 179& 16.33& 0.0& 0&
\\
2861& 7493& -367&   378& 16.26&  0.00& 1.85& 0.2& 134& 15.02& 2.0& 0&
\\
2914& 7560& -313&   680& 17.18& 15.40& 1.81& 0.2& 117& 15.89& 0.0& 0&
\\
3084& 7566& -132&   568& 16.43&  0.00& 1.82& 0.1&  39& 15.30& 1.0& 0&
\\
3113& 7546&  -87&   461& 17.82& 16.08& 1.81& 0.1& 147& 16.65& 0.0& 0&
\\
3254& 7512&   65&    -8& 16.57&  0.00& 1.84& 0.4&  55& 15.87& 2.0& 2&
\\
3486& 7604&  263&  -130& 17.73& 15.79& 1.82& 0.1& 136& 16.67& 1.0& 0&
\\
3487& 7601&  263&    -9& 16.63&  0.00& 1.88& 0.4&  67& 15.80& 2.0& 0&
\\
3640& 7483&  398&  1011& 17.13& 16.23& 0.00& 0.4&  95& 16.41& 0.0& 0&
\\
3761& 7650&  519&    96& 15.57&  0.00& 1.88& 0.3&  51& 14.43& 2.1& 0&
IC3955\\
\hline
\end{tabular}
\end{center}
}}
\end{table}

\begin{table}                 
\caption{Subgroup 3.}
{
\renewcommand{\baselinestretch}{1.2}
\renewcommand{\tabcolsep}{.95mm}
{\footnotesize
\begin{center}
\begin{tabular}{ccrrcrccrcccl}
\hline
\hline\\[-3.5mm]
GMP& VEL& \multicolumn{1}{c}{X}& \multicolumn{1}{c}{\,\,\,\,Y}& B\,&
\multicolumn{1}{c}{\,\,R}& B-R & Ell & The & V\,& Type & Class & Ident\\
\hline\\[-3mm]
2921& 6497&  -304&    22& 12.62&  0.00& 1.91& 0.0&   0& 11.78& 1.0& 0&
NGC4889\\
2252& 5979& -1114&  -548& 16.10&  0.00& 1.85& 0.3&  56& 15.08& 1.0& 0&
\\
2516& 6255&  -762&     3& 15.34&  0.00& 1.86& 0.0&   0& 14.21& 2.5& 0&
IC4042 \\
2805& 6141&  -422&   337& 16.57&  0.00& 1.78& 0.4&  73& 15.47& 2.0& 0&
\\
2945& 6191&  -278&  -703& 16.15&  0.00& 1.77& 0.6& 148& 15.02& 2.0& 0&
\\
2960& 5922&  -267&   194& 16.78&  0.00& 1.74& 0.4& 159& 15.79& 2.0& 0&
\\
3205& 6200&    14&  -371& 17.61& 15.76& 1.83& 0.3& 138& 16.38& 0.0& 0&
\\
3313& 6210&   111&  -521& 17.53&  0.0 & 1.83& 0.1& 151&  0.00& 0.0& 0&
\\
3367& 5848&   166&    47& 15.15&  0.00& 1.91& 0.3& 108& 14.42& 2.0& 0&
NGC4873\\
3484& 6082&   262&     9& 16.26&  0.00& 1.81& 0.4&  50& 15.49& 2.0& 0&
\\
3493& 6008&   270&  -833& 16.50&  0.00& 1.94& 0.5&   9& 15.22& 2.0& 0&
\\
3879& 5967&   630& -1352& 16.31&  0.00& 1.86& 0.3& 101& 15.08& 2.0& 0&
\\
3972& 6018&   721&   410& 16.52&  0.00& 1.87& 0.6& 167& 15.29& 2.0& 0&
\\
3997& 5983&   751&  -575& 15.28&  0.00& 1.95& 0.0&   0& 14.15& 2.0& 0&
IC3946 \\
4083& 6202&   856&  -516& 17.82& 15.84& 1.91& 0.4&  79& 16.44& 2.0& 0&
\\
4103& 5978&   878&   -59& 17.74& 15.96& 1.76& 0.6& 126& 16.21& 0.0& 0&
\\
4315& 5994&  1104&    -6& 15.39&  0.00& 1.87& 0.0&   0& 14.20& 1.5& 0&
NGC4850\\
\hline
\end{tabular}
\end{center}
}}
\end{table}

\begin{table}
\centering
\caption{Parameters of the Coma core main system (MS) and
subgroups (1s,
2s, 3s): N denotes the total number
of galaxies in the initial sample (T) and in each system;
$m$ the median
velocity; $\sigma, s, c$, the standard deviation, 3rd and
4th moment of
redshift distribution, respectively.}
\medskip
{\footnotesize
\begin{tabular}{ccrrrr}
\hline
\hline
Coma core& T ($<18m$)& MS    &   1s &  2s  & 3s  \\
\hline
N        & 188       &   174 &   34 &   14 &  17 \\
m        &           &  6953 & 6892 & 7563 & 6013\\
$\sigma$ &           &   949 &  206 &   60 &  122\\
s        &           &  -0.2 &  0.4 &  0.4 &  0.1\\
c        &           & -0.86 & -1.1 & -1.0 & -1.4\\
\hline
\end{tabular}}
\end{table}

\newpage

{\it Figure captions}.

Figure 1. The redshift histogram of the main system (MS) of the
core of Coma cluster.
The revealed three galaxy subgroups are indicated
(dashed lines), with the
parameters given in Table 1.

Figure 2. The histogram of orientations of the galaxies in subgroup 1.

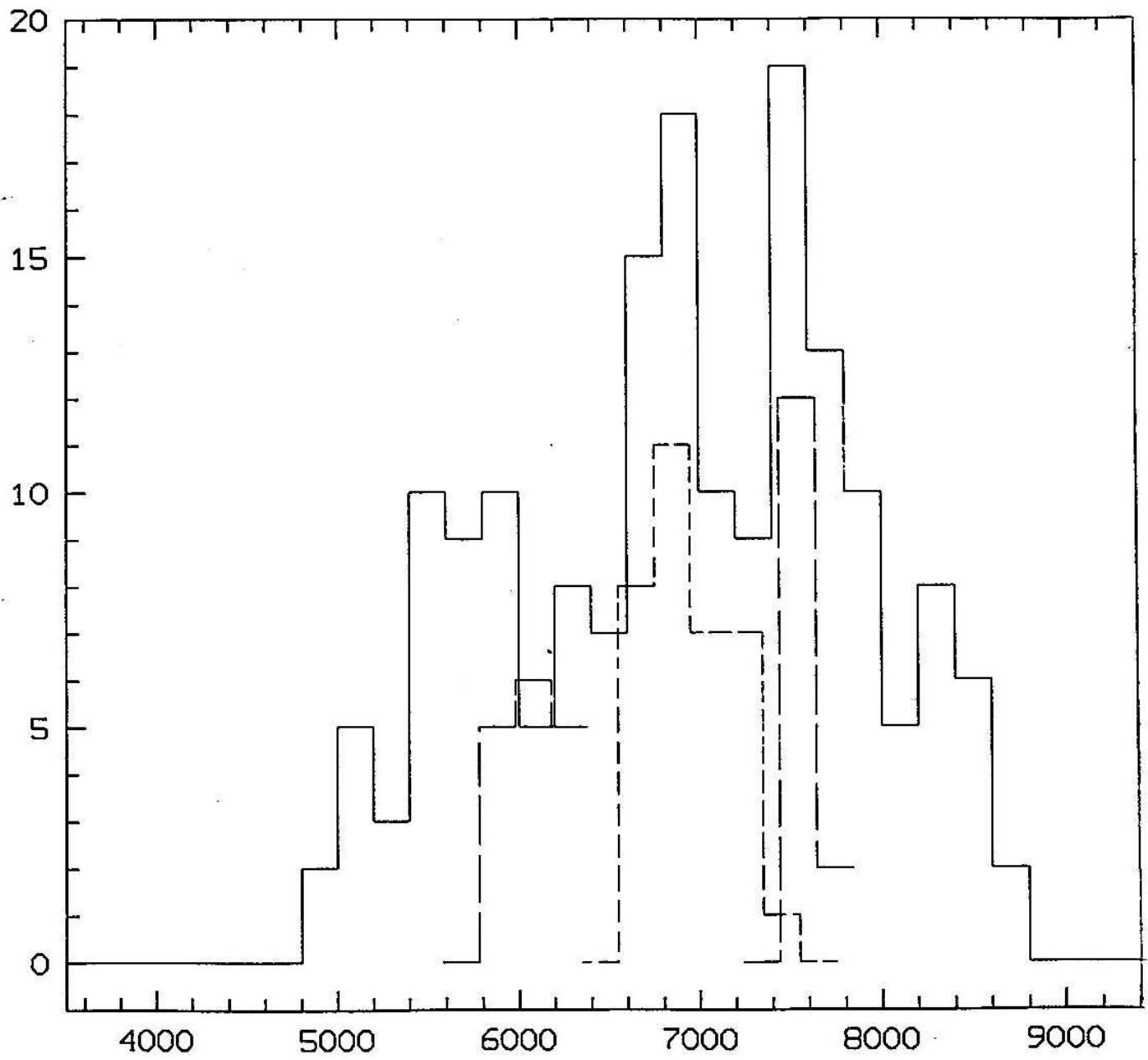
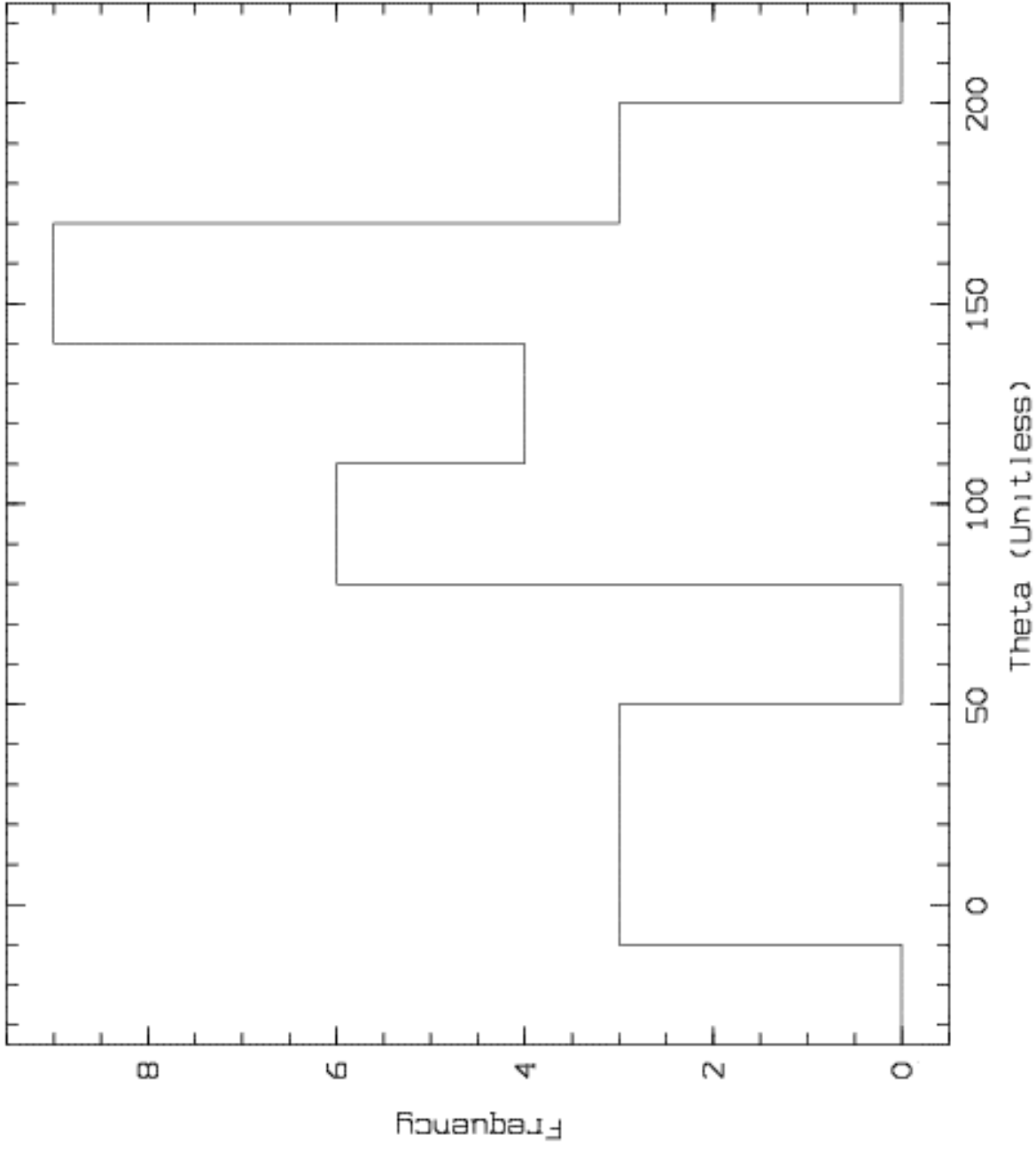
\end{document}